# HydroMind: A Checkable Agent for Autonomous Hydrological Research

Zheng Fang[1], Yiqun Sun[1], Simin Qu[1], Shijie Jiang[2]

[1]College of Hydrology and Water Resources, Hohai University, Nanjing, China

[2]Department of Biogeochemical Integration, Max Planck Institute for Biogeochemistry, Jena, Germany

## Abstract

General-purpose coding agents have written code for hydrological and Earth-system models. In these studies the researchers decided in advance how the model would be evaluated, for example the test period, the scoring function, and the standard the model had to meet, and the agent wrote code to meet them. No study has tested what happens when the agent makes these decisions itself: how many of its claims are supported by its own data, code, and results, and whether readers can check them. Both may depend on the agent's harness, that is, the tools it can call and the written instructions it follows, which its user sets. We compared three harnesses. The first is the agent's own harness, with nothing specific to hydrology. The second adds the modeling tools of HydroMind, a hydrological modeling system we developed for general-purpose agents. These tools include 84 hydrological models and common loss functions and metrics, and the agent can add its own models and use them in the same way. Training, prediction, and metric computation run as fixed commands that store the settings, trained model, and predictions of every run, so each reported number can be traced to the run that produced it. The third harness also adds HydroMind's research procedure: the agent writes a research design before running any model and records every change to it. Under each harness, the same agent studied four open questions in hydrological model development and was given only the data. Each experiment submitted a list of findings. Findings that state one model or approach is better than, worse than, or similar to another are claims; the others describe the data, methods, or scope. The experiments under the three harnesses submitted 23, 20, and 41 claims. We checked each claim in two ways. We recomputed the stated difference from the experiment's own code and predictions and compared it with the claim, and we read the code to see whether it compared the two models or approaches the claim names. A claim that fails either check does not hold. Under the three harnesses, 30%, 20%, and 5% of the claims did not hold. Readers can judge a stated difference only if the experiment reports a statistical test, a confidence interval, or a decision rule, such as how large a difference counts as an improvement. Such support was reported for 48%, 53%, and 83% of the claims. With the language model unchanged, changing only the harness reduced the share of claims that do not hold and increased the share that readers can check.

## 1 Introduction

General-purpose coding agents, agents based on large language models that write and run code and are not specialized for any specific domain, have written code for hydrological and Earth-system models. Earlier hydrological agents were mostly used to execute specified modeling tasks: operating modeling software (Ren et al., 2024; Pursnani et al., 2024; Tudaji et al., 2026), and calibrating and running existing models (Li et al., 2026a, 2026b; Ma et al., 2026a). Recently, agents wrote most of the code for a differentiable Earth-system model with soil-water and soil-heat components (Gentine et al., 2026), translated a 19,000-line Fortran land surface model into a differentiable framework (Lahlou et al., 2026), explored model architectures for seasonal streamflow forecasting in the Sierra Nevada (Lopez-Gomez et al., 2026), and proposed architectures for bucket-type conceptual models (Adombi, 2026). These studies share a common feature: how the agent's results were evaluated was determined in advance by researchers. Conventions and acceptance criteria for the Earth-system model were specified by a written scientific contract (Gentine et al., 2026), and the forecasting study fixed the task, test period, and scoring function (Lopez-Gomez et al., 2026). The agent's task was only to write code that satisfied these evaluations.

In hydrological modeling studies, research decisions (such as the choice of training, validation, and test periods, loss function, study area, and evaluation criteria) often change the conclusions. Existing studies show that water resources assessment results for the same set of basins vary with the chosen models and parameters (Melsen et al., 2018). Even for the same basin, different modelers produced widely differing predictions, because each modeler made different research decisions (Holländer et al., 2009, 2014). In some cases, changing only the evaluation metric altered conclusions about flood simulation (Melsen et al., 2019). The choice of model often depends on the modeler's prior experience rather than on the suitability of the model (Addor & Melsen, 2019). The same holds when agents perform data analysis: multiple agents given the same data and research question reach different conclusions (Bertran et al., 2026; Gao & Xiao, 2026; Miao et al., 2026). Agents have also carried a study from an open research question through to its conclusions (Lu et al., 2026; Pursnani et al., 2026). In such studies, research decisions are made by the agent, and its conclusions therefore depend on decisions that researchers did not fix in advance.

When an agent carries a study from an open question to claims, there is no reference answer to evaluate the claims against, because most research decisions have more than one reasonable choice. Different choices can yield different conclusions; what can be tested is whether the study is internally consistent. For example, every claim stated by a study should be supported by relevant records, such as the analyzed data, code, run results, and statistical analyses. The agent's report alone is not sufficient to perform this check; execution logs of research agents often show failures not reflected in their final papers (Luo et al., 2025). Across 100 open-ended scientific analysis cases, no agent system fully matched claims to evidence in more than 18.0% of cases (Liu et al., 2026); review panels looking only at code and scores missed 6.5% of confirmed cases where research agents manipulated evaluations (Huang et al., 2026). Therefore, checking claims requires research records in addition to the report: which models were run, what the settings and results were, and what decisions the agent made and when. When agents conduct hydrological model development research, it has not been tested how many claims hold and whether readers can check them against the records the research leaves.

Which records a study leaves may depend on the agent's harness (Zhang & Valeo, 2026; Xia & Wang, 2026). The harness is the part of an agent outside the language model that is set by the user, namely the tools the agent can call and the written instructions it must follow. Existing studies disagree on whether changing the harness changes an agent's practices in research. Across a large volume of agent execution traces, the harness explained agent behavior to a much lower degree than the language model did (Ríos-García et al., 2026). In other studies, specialized verification tools changed agent behavior while verification prompts did not (Wiedmann et al., 2026); a research procedure that accepts only claims with recorded evidence reduced the share of unsupported claims from approximately 13.3% to 4.9% (Xia & Wang, 2026). The choice of harness also changed evaluation results of coding agents (Vats &

Golev, 2026), and the evaluation protocol provided to an agent changed the validity of the models it constructed (Ma et al., 2026b). None of these studies was in hydrology. In hydrological research where the agent makes the research decisions itself, it has not been tested whether an agent can conduct research autonomously, whether such research is credible, or whether the harness can improve its credibility.

To this end, we developed HydroMind, a hydrological modeling system for general-purpose agents, comprising modeling tools and a research procedure. The modeling tools provide built-in hydrological models, loss functions, optimizers, and evaluation metrics, together with fixed commands in a command-line interface (CLI) for reading basin data, training, predicting, and computing metrics; for every model run made with these commands, the configuration (such as the hyperparameters) and the metrics during training are recorded automatically. HydroMind's research procedure requires the agent to determine research decisions item by item and write them into a research design document, run model experiments in rounds, and record relevant findings at the end of each round. To test HydroMind, we compared three harnesses. The first is the general harness, that of Codex: its tools edit files and run commands, and it contains nothing specific to hydrology. The second is the general harness plus the HydroMind modeling tools, and the third is the general harness plus the HydroMind modeling tools and the research procedure. We prepared four research questions to test each harness: how to improve the simulation of a specific hydrological model in basins not used in training; how to represent reservoir release in a model; how the number and composition of training basins affect a regional LSTM; and whether recent streamflow observations are better used as model inputs or to update model states. For each question, the agent received only the data and made all research decisions itself. We checked every claim submitted under each harness, one by one, to determine whether the experiment's own materials support it. The check has two parts: first, we recomputed the difference stated in the claim from the code and final predictions submitted by the agent and checked whether it agrees with the claim; second, we read the documentation and code and checked whether the code actually compared what the agent claimed. A claim that fails either part does not hold. Claims that do not hold made up 30% (7/23), 20% (4/20), and 5% (2/41) under the three harnesses. We also counted how many claims were accompanied by statistical tests, confidence intervals, or decision rules provided by the experiments themselves (such as whether the NSE difference is robust across different hyperparameters). Under the three harnesses, 48%, 53%, and 83% of the claims came with such support.

## 2 HydroMind

HydroMind is a hydrological modeling harness we developed for general-purpose agents, comprising modeling tools and a research procedure. Like other harnesses, HydroMind provides the agent with callable tools and written instructions. HydroMind has two layers: the modeling tools and the research procedure. The modeling tools are a comprehensive hydrological modeling toolkit, described in detail in Section 2.1. The research procedure is an orchestrator built on top of the modeling tools, and abstracts the process of hydrological model research. It is an abstract research methodology rather than a specific workflow.

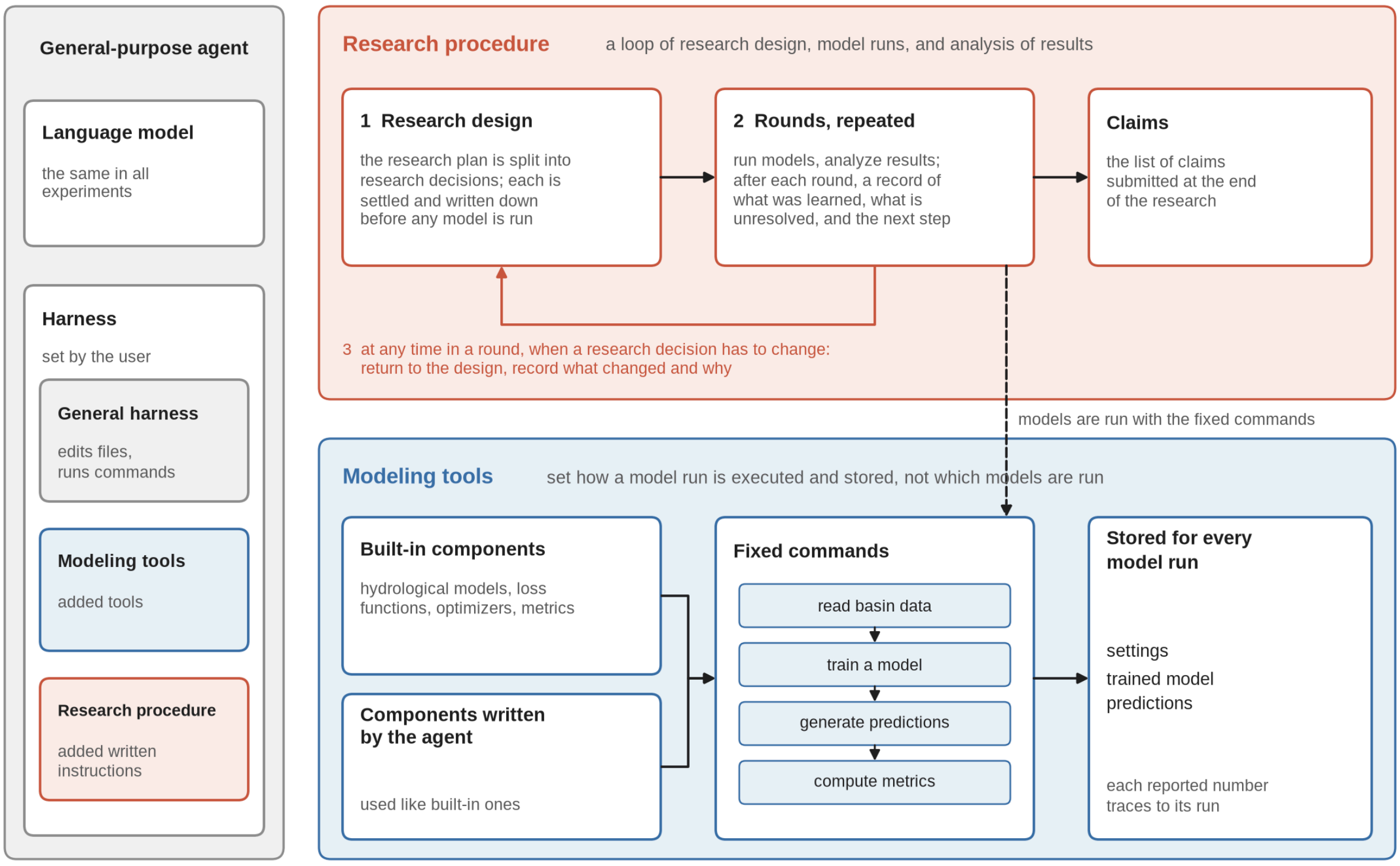

**Figure 1.** The two parts of HydroMind and their place in a general-purpose agent. Left: an agent consists of a language model and a harness; HydroMind adds the modeling tools and the research procedure to the harness. Top right: the research procedure. The agent writes a research design and then runs models and analyzes results round after round, with a record at the end of each round; at any time, when a research decision has to change, the agent returns to the design, changes it, and then continues. Bottom right: the modeling tools. Built-in components and components written by the agent are used through the same fixed commands, and the settings, trained model, and predictions of every model run are stored.

## 2.1 Modeling Tools

With the general harness alone, an agent that studies a hydrological modeling question writes all the code itself: reading the data, building the model, training, predicting, and computing metrics. Which script and which settings produced a result is known only if the agent itself wrote it down. The modeling tools turn these steps, which every modeling study contains, into fixed commands, and they store every model run automatically.

The modeling tools form a Python package with a command-line interface and a front-end interface. The agent operates the package through Agent Skills and the CLI. Training a model with them takes three steps. First, the agent registers the data: it puts the meteorological, streamflow, and attribute data of the basins into one file and registers this file once, and every later model run refers to these data by name. Second, the agent writes the settings. They state the model and its hyperparameters, the basins, the input and target variables, the training, validation, and test periods, the loss function, the optimizer, the metrics, and the random seed. Settings cannot be modified once they are created; to change any item, the agent creates new settings. Third, the agent submits the settings with one command, and the modeling tools train the model and evaluate it with the metrics named in the settings. Generating predictions for another period with a trained model is again one set of settings and one submission. The agent writes none of the code that reads the data, trains the model, or computes the metrics.

The modeling tools have 84 built-in models. Of these, 31 are long short-term memory (LSTM) networks, gated recurrent units (GRU), and their variants; 33 are Transformer-type models and other time-series networks (such as Informer, PatchTST, and Mamba); 11 are MC-LSTM and its variants; and 9 are differentiable conceptual models (HBV, GR4J, SAC-SMA, and the Xinanjiang model). It also has 22 built-in evaluation metrics (such as the Nash–Sutcliffe efficiency and the Kling–Gupta efficiency), 16 loss functions, 15 model optimizers, and 14 learning-rate adjustment methods.

When a study needs a model that is not among these 84, the agent can add its own model through HydroMind's registration interface. For example, for the question in Section 3 regarding how to use recently observed streamflow to improve model performance, the agent created a Python code file for an LSTM that corrects internal states with streamflow observations and registered it using the HydroMind model registry. This model was then trained and evaluated with the same commands as the built-in models. The agent can register new evaluation metrics, loss functions, and optimizers in the same way.

Each time a model is trained, validated, or tested, the modeling tools automatically save three items: all parameter settings for that run (such as all hyperparameters), the trained model checkpoint, and the model's predictions and the metrics recorded during training. The agent does not need to keep separate records. Therefore, when the agent reports a metric value for a model, this number can be traced directly to a record in the database, and from there to the settings and code of that run. Researchers checking the agent's claims can inspect HydroMind's internal database directly for the details of training, validation, or testing. HydroMind also provides a visual interface and a CLI that the agent can use to inspect records.

## 2.2 Research Procedure

The research procedure is a set of written instructions that divides a study into three stages: research design, running models, and analyzing results, specifying what records each stage leaves. The three stages are not executed once in sequence, but iteratively in a loop (Figure 1). The agent first writes the research design, and then runs models and analyzes results round by round, with each round planned according to the record of the previous round. In any round, when a decision in the research design may need to change, the agent decides, according to the rules of the research procedure (see below), whether to return to the design stage, and continues once the change is made. The loop continues until the agent judges that the research question has been answered. The research procedure does not prescribe which metrics to use, what to compare, or what claims to draw.

The loop starts with the research design. Before any model is run, the agent divides the research plan into individual research decisions, asks a question about each, and gives a recommended answer, until every decision has concrete content, and then writes the research design. For example, one research question in Section 3 asks whether recent streamflow observations are better used as a model input or to update the model states. When the agent studied this question under the harness with the research procedure, the research design that it wrote settled the following. All 531 basins are used. The years 1999–2008 are used for training, 1980–1989 for model selection, and 1989–1999 for the final evaluation. Three models are compared: an LSTM with meteorological inputs only, an LSTM that receives recent streamflow as an input, and an LSTM whose states are corrected with recent streamflow; the streamflow observations have a delay of 1 or 5 days. The main quantity is the median of the differences in Nash–Sutcliffe efficiency between the latter two models, paired by basin. The requirement that this difference has to meet before one way is stated to be better was also written down at this time (Table 1). In normal use a human researcher takes part in the discussion of this step. In the experiments of this paper the agent settled every decision itself.

After the research design is written, the agent works in rounds. A round serves one purpose, for example to compare the three models in the research design or to find the cause of a result, and it can contain many model runs. In a round the agent runs models and then analyzes the results. At the end of the round it writes a record that states what was learned in the round, what remains unresolved, and what is to be done in the next round. The next round is planned from this record, and the rounds repeat in this way.

Returning to the design does not have to wait for the end of a round; it can happen at any time while running models or analyzing results. The research procedure gives rules for this judgment: if a choice at hand would change the research question, the meaning of a comparison, the meaning of the data, or the interpretation of results, the agent stops the work at hand and returns to the design; other choices, such as adjusting hyperparameters or correcting code errors, are decided directly by the agent without returning to the design. After returning to the design, the agent settles this decision again using the method for writing the research design, revises the research design, writes down what was changed and why, and then returns to the original round to continue. In the study of "recent streamflow as an input or for correcting model states" described above, partway through the first batch of training, one model failed due to insufficient GPU memory. Reducing the batch size for this model alone would make its training differ from that of the other models and change the meaning of the comparison; the agent therefore returned to the design and wrote that the batch size was halved for all models, that all nine training runs were repeated, and that the models trained before were no longer used, because models trained with different batch sizes cannot be compared. The research design, the records of each round, and the records of changes are the other part of the records mentioned in Section 1: what decisions the agent made and when.

## 3 Experimental Design

To test whether the harness changes the share of a general-purpose agent's claims that hold and whether readers can check these claims, we designed a controlled experiment: the same agent studied the same four hydrological modeling questions under three harnesses, yielding 12 experiments (Figure 2). Below we describe the four research questions, the three harnesses, and how the claims submitted by each experiment were evaluated.

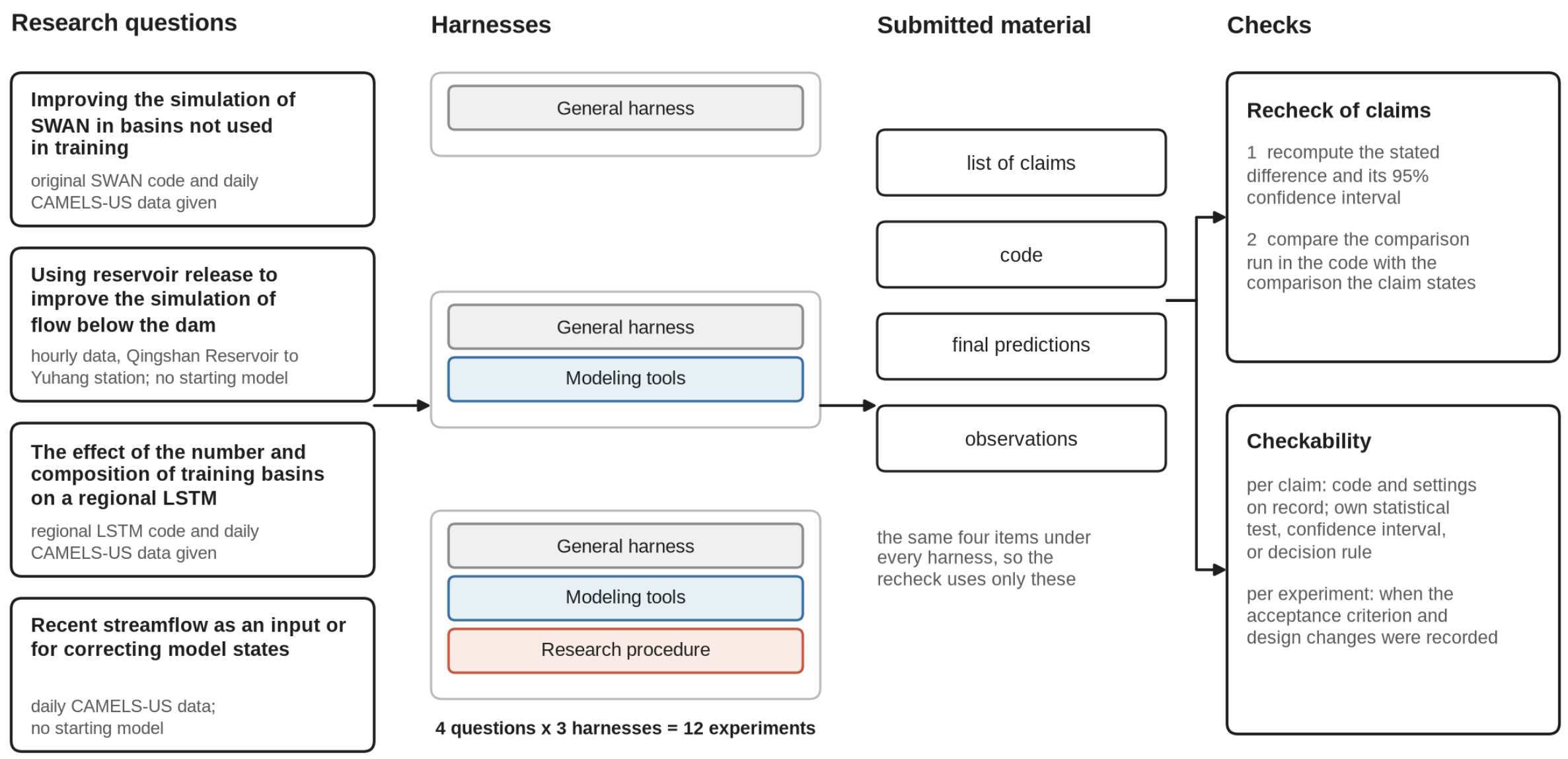

**Figure 2.** Experimental design. The same agent studied four research questions under three harnesses, yielding 12 experiments. Each harness adds one part to the previous one. Each experiment submitted the same four items. We rechecked claims from the submitted materials and assessed what records each experiment provided for checking its claims.

The four research questions represent types common in hydrological model development: two require improving a simulation, and two require evaluating the effect of a practice or a factor. For each type, two questions with different study subjects were selected so that we can check whether a difference between harnesses appears in both questions of a type rather than in only one question. Each question states only what is to be answered and not how to answer it. All observation data were open to the agent, and the agent itself partitioned which data to use for training and which for final evaluation.

The first question is: how to improve the simulation of a specific model in basins not used in training. It represents a basic category of work in hydrological model development: proposing an improvement for a deficiency in an existing model and testing whether the improvement holds. Hydrological models often need to be applied to basins without streamflow observations, where model performance is typically lower than in training basins; improving model simulation in these basins is therefore a common research goal. The difficulty in this type of research is that an average improvement across all basins can mask degradation in certain basins or across certain flow ranges. The agent received the original code of SWAN (Fang et al., 2025) and daily CAMELS-US data (Newman et al., 2015; Addor et al., 2017), and could modify the model structure or training method. SWAN is a neural network hydrological model with water mass accounting, and its performance varies markedly across different basins (Fang et al., 2025). The question required the claims to answer: whether the modified method is superior to the original SWAN on evaluation basins designated by the agent itself; in which basins improvements and degradations occur; and whether the effects on low, medium, and high flows are consistent.

The second question is: how to use information on reservoir release in a hydrological model to improve the simulation of streamflow downstream of a dam, while maintaining a hydrological interpretation and water balance that can be checked. It represents another situation in model improvement: the model must represent human-regulated processes. Reservoirs alter runoff regimes in many basins. Large-scale studies show that reservoir purpose and degree of regulation clearly affect model performance (Ouyang et al., 2021); recent studies have compared different ways of coupling reservoir operation modules with rainfall-runoff modules (Chen et al., 2025). Unlike the first question, improving simulation accuracy alone is not sufficient here. The study area is the 114 km² area between Qingshan Reservoir and the downstream Yuhang station, where streamflow at Yuhang station is formed jointly by reservoir release and runoff generated within this area. The agent received hourly data for this area, in which observed rainfall and release 1–7 h ahead were given as known inputs. No starting model was provided; which model to use and which baselines to compare against were chosen by the agent.

The third question is: how the number and composition of training basins affect the simulation of regional LSTM models. The third and fourth questions belong to the second type, which asks for the effect of a practice or a factor. Both questions have been answered in published studies; we did not provide the procedure or conclusions of these studies, and the agent designed the comparisons and answered on its own. Regional models are trained jointly on data from many basins, and how many basins and which basins to use are choices modelers must make. This choice is made in many studies: among 150 LSTM rainfall-runoff papers from 2021–2023 surveyed by Kratzert et al. (2024), 122 were still trained on individual basins. Kratzert et al. (2024) studied this question by systematically varying the number and composition of basins trained together. The difficulty in this type of research is separating the effect of basin count from that of basin composition. The agent received common regional LSTM code and daily CAMELS-US data. The question required the claims to answer: whether training on more basins together improves simulation on the same evaluation basins in an independent period; whether the hydrological composition of basins changes the results when the number of basins is comparable; and whether observed differences can be separated from the effects of training sample size and training computation.

The fourth question is: for an LSTM driven by meteorological data, how to use newly observed streamflow to improve the estimate of current flow: whether it is better to provide recent streamflow directly as a model input or to use it to update the internal states of the model. It represents questions about how to use observations. Recent streamflow observations are commonly used information in operational forecasting; Nearing et al. (2022) summarized direct input and data assimilation as two primary pathways for using recent streamflow, and compared them on CAMELS data. The difficulty in this type of research is that the comparison of the two approaches has to take the delay of observations and missing observations into account. As with the third question, the agent did not receive the procedure or conclusions of the original study. The agent received daily CAMELS-US data. No starting model was provided; both the LSTM baseline and the implementations of the two approaches were completed by the agent.

For each question the agent received only the data and code named above, and it made all research decisions itself. The questions did not prescribe which basins and periods to use for the evaluation, what to compare with what, or which metric to treat as the main one, and they gave no acceptance criterion, that is, no requirement that a result must meet before it is stated as a claim, such as how large a difference counts as an improvement. The agent also decided when the research was complete. The questions were given in this form because this study examines whether the claims of an agent hold when the agent makes the research decisions itself (Section 1).

The same agent studied each of the four questions under three harnesses, yielding 12 experiments; one experiment is one research question studied by the agent under one harness, from the question to the claims. The agent was Codex with the language model GPT-5.6 Sol at medium reasoning effort. Of the three harnesses, the first is the general harness, which is the harness of Codex as supplied. Its tools edit files and run commands, and it contains nothing specific to hydrology; under this harness, the agent worked in a standard scientific Python environment. The second is the general harness plus the HydroMind modeling tools. The third is the general harness plus the HydroMind modeling tools and the research procedure. The language model, computing resources (one NVIDIA RTX 5090 D graphics processing unit), and question text were the same under the three harnesses, and each experiment started from a new context. Each experiment submitted the same four items: a list of findings, code, final predictions, and experimental records.

This paper addresses two questions: how many of the agent's claims hold, and whether readers can check these claims against the records left by the experiments. The four research questions differ in content, but the agent's claims have the same form: over specified basins and periods, according to a given metric, one model or practice is superior to, inferior to, or close to another. For example, one claim in the question "using reservoir release to improve the simulation of flow below the dam" is that the mean absolute error of the new model during high-flow periods is lower than that of the benchmark model (0.287 vs. 0.320 mm/h). Such a claim makes two statements: what was compared, and how large and how certain the resulting difference is. A claim holds only when both statements are supported by the experiment's own materials. We therefore check each claim in two ways. The first check recomputes the stated difference from the results submitted by the experiment and judges, against the wording of the claim, whether the difference holds. The second check reads the code to confirm that the comparison it actually runs is the comparison the claim states; if the code compares other objects, the claim does not hold even if the difference can be reproduced. Other differences in setup between the two groups do not count as a failure if they are stated in the claim or arise from the compared factor itself. A claim that fails either check is recorded as not holding.

Readers checking a study conducted by an agent can rely only on the records left by the study itself. We therefore also assessed the records left by each experiment. For each claim, we assessed two items: whether the code and settings producing the numbers in the claim are preserved in the version actually run; and whether the experiment provides its own statistical test, confidence interval, or decision rule for that comparison. In addition, for each experiment, we examined when research decisions were written down, looking specifically at two points. First, whether the agent wrote down, before seeing the final results, how large a difference counts as an improvement, that is, the acceptance criteria. If acceptance criteria are determined after seeing the results, readers cannot tell whether the results met the criteria or whether the criteria were tailored to the results. Second, whenever the agent modified the research design, whether it wrote down at the time what was changed and why. With such records, readers can see when and why the design changed.

# 4 Results

## 4.1 Share of Claims That Do Not Hold

The claims that do not hold make up 30% (7 of 23), 20% (4 of 20), and 5% (2 of 41) under the three harnesses (Figure 3a). The lists of findings submitted by the 12 experiments contain 118 items in total, 84 of which are claims: 23 under the general harness, 20 under the general harness with the modeling tools, and 41 under the general harness with the HydroMind modeling tools and the research procedure.

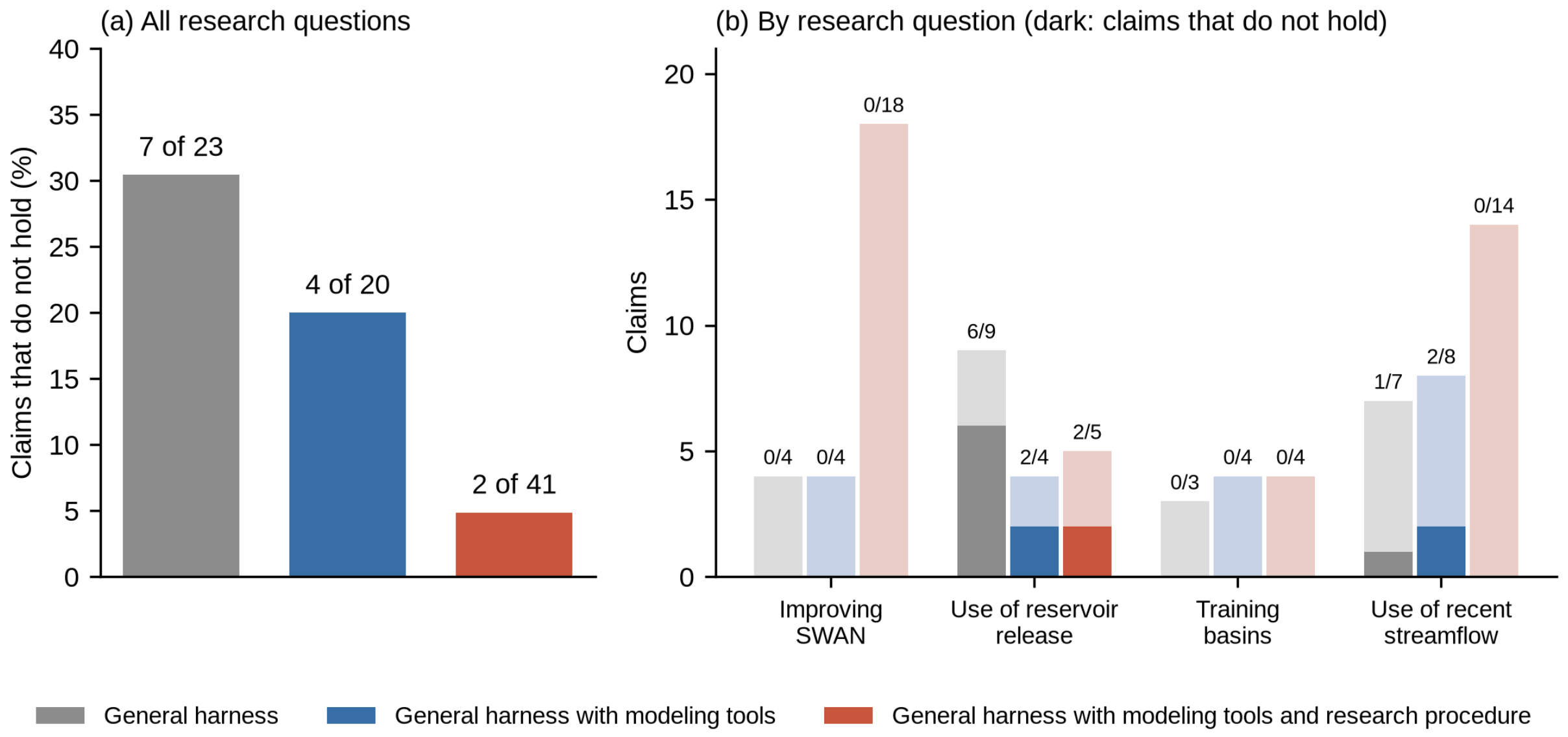

**Figure 3.** Claims that do not hold. (a) Share of the claims that do not hold, under each harness. (b) The same claims by research question: the light bar is the number of claims, the dark bar the number that do not hold.

The claims that do not hold are concentrated in the two questions "using reservoir release to improve the simulation of flow below the dam" and "recent streamflow as an input or for correcting model states" (Figure 3b). On the former question, the numbers under the three harnesses are 6 of 9, 2 of 4, and 2 of 5; on the latter, 1 of 7, 2 of 8, and 0 of 14. On the two questions "improving the simulation of SWAN in basins not used in training" and "the effect of the number and composition of training basins on a regional LSTM", there are no claims that do not hold under any of the three harnesses.

Across the three harnesses, a total of 13 claims do not hold (7 under the general harness, 4 with the modeling tools added, and 2 with the research procedure added further, namely the three numbers in Figure 3a). All of them are of the same kind: the claim states that one model or practice is better or worse than another, but the difference cannot be distinguished from random variation. For example, one claim states a reduction in mean absolute error of 0.033 mm/h, but the recomputation shows that this difference lies within statistical uncertainty. Among these 13 claims, only 1 has an additional problem: reading the code reveals that it actually compared groups other than the two stated in the claim. This claim comes from the experiment on "using reservoir release to improve the simulation of flow below the dam" under the general harness. For 10 of the 13 claims (5 of 7, 3 of 4, and 2 of 2 under the three harnesses), the experiment provided no statistical test, confidence interval, or decision rule of its own for the stated difference, and the claim rested only on the value of the difference itself (Section 4.2).

### 4.2 Checkability of Claims

The setup of the comparison is on record for 89%, 100%, and 100% of the claims under the three harnesses (Figure 4a). Section 4.1 reported the results of our recheck; this section reports the records provided by the experiments themselves. With the modeling tools, every claim has this record (20 of 20 and 41 of 41). Under the general harness, the share is 20.5 of 23; the missing records are in the experiment on “recent streamflow as an input or for correcting model states”: the training script was changed after part of the training had started, so for five claims the script on file cannot be confirmed to be the version actually run, and each of these five counts as one half.

The claims whose stated difference is supported by the experiment's own statistical test, confidence interval, or decision rule make up 48%, 53%, and 83% (Figure 4b). The denominator leaves out claims that compare only computing time, which are 21, 19, and 41 under the three harnesses. The modeling tools had little effect on this share (10 of 21 and 10 of 19), and the research procedure raised it to 83% (34 of 41). In the question “recent streamflow as an input or for correcting model states”, the experiments under the first two harnesses attached no test or interval to any comparison (0 of 6 and 0 of 7), while the experiment with the research procedure attached significance tests to its comparisons (13 of 14).

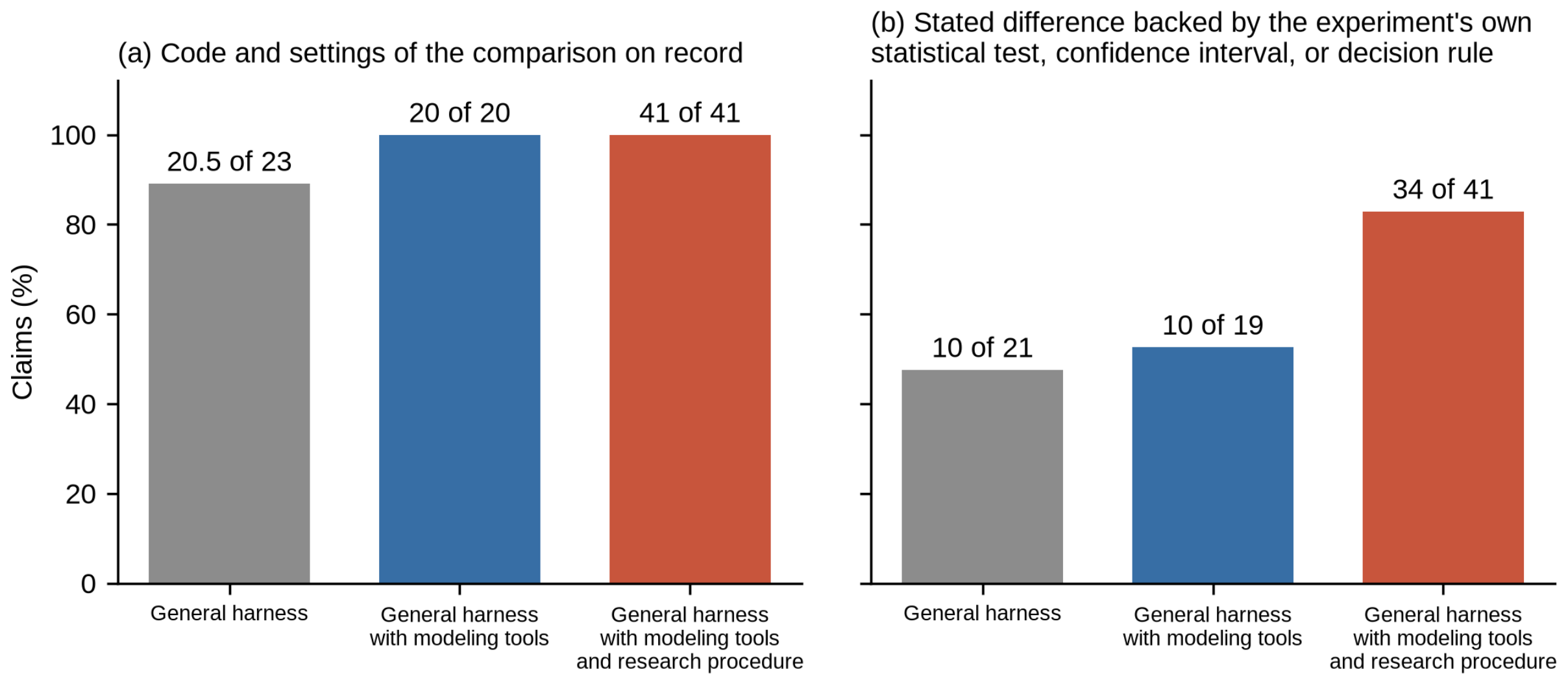

**Figure 4.** Records for checking the claims, under each harness. (a) Share of the claims for which the code and settings of the comparison are on record. (b) Share of the claims, without those that compare only computing time, whose stated difference is supported by the experiment's own statistical test, confidence interval, or decision rule.

Only experiments with the research procedure wrote down acceptance criteria before seeing the final results: the experiments on all four research questions did so (Table 1). None of the eight experiments under the general harness or the general harness with the modeling tools wrote down acceptance criteria beforehand. The records of changes to the research design show the same difference. In the four experiments with the research procedure, every change to the design was recorded at the time, with what was changed and why; under the other two harnesses, only some of the changes were recorded at the time. For example, in the experiment on “improving the simulation of SWAN in basins not used in training” under the general harness with the modeling tools and the research procedure, the agent found that a new training setup reduced the validation loss, but the training period of this setup overlapped the validation period, making the validation results unreliable. The agent recorded at the time the decision to discard this result and the reason for it, separated the training and validation periods again, and then continued the research.

**Table 1.** Acceptance criteria that the agent wrote down before it saw the final results, that is, the requirements a result had to meet before it was stated as a claim. Only the experiments under the general harness with the HydroMind modeling tools and the research procedure are listed; no experiment under the other two harnesses wrote down acceptance criteria.

| Research question | Acceptance criterion |
|---|---|
| improving the simulation of SWAN in basins not used in training | The modified model is significantly superior to the original SWAN across all 10 sets of hyperparameters and 3 random seeds |
| using reservoir release to improve the simulation of flow below the dam | Closure of the water balance: 99th percentile of the relative residual below $10^{-5}$ and maximum below $10^{-4}$ |
| the effect of the number and composition of training basins on a regional LSTM | Median paired difference in Nash–Sutcliffe efficiency (NSE) of at least 0.02; difference in volume bias at high flows of at least 5 percentage points, statistically significant |
| recent streamflow as an input or for correcting model states | Median paired NSE difference of at least 0.01, statistically significant; agreement in the Kling–Gupta efficiency results; degradation in missed peaks of no more than 0.02 |

## 5 Discussion

With the language model held constant, changing only the harness changes the share of claims that do not hold and the share of claims accompanied by the experiment's own statistical support. Under the general harness, claims that do not hold accounted for 30%, and the share of claims with statistical support was 48%; under the general harness plus the HydroMind modeling tools, claims that do not hold accounted for 20%, and the share with statistical support was 53%; and under the general harness plus the HydroMind modeling tools and the research procedure, claims that do not hold fell to 5%, and the share with statistical support reached 83% (Sections 4.1 and 4.2). All 13 claims that do not hold are of the same kind: the claim stated that one model or method was better or worse than another, but the difference could not be distinguished from random variation. Only one of the 13 also compared, in its code, something other than what the claim states; this claim appeared in the experiment on “using reservoir release to improve the simulation of flow below the dam” under the general harness. Ten of the 13 claims lacked the experiment's own statistical test, confidence interval, or decision rule. In these experiments, therefore, the main problem was that differences not shown to exceed random variation were reported as claims; coding errors accounted for very few of the failures.

The general harness plus the HydroMind modeling tools changed what the experiments recorded. Under this harness, the code and settings behind every claim were on record (reaching 100%, compared with 89% under the general harness), and the only claim where the comparison in code did not match the statement appeared under the general harness (Sections 4.1 and 4.2). The share of claims that do not hold fell from 30% to 20%, but the share of claims with statistical support changed very little (48% and 53%, respectively), and all four claims that did not hold under this harness belonged to the same type described above, three of which had no statistical support from the experiment itself. This means that fixed commands and the automatic saving of every model run make it possible to trace what was run, but they do not determine whether a difference is large enough or certain enough to be reported as a claim.

The general harness plus the HydroMind modeling tools and the research procedure changed how claims were judged. With the research procedure added, the share of claims with statistical support rose from 53% to 83%, and the share of claims that do not hold fell from 20% to 5% (Sections 4.1 and 4.2). In all four experiments, the agent wrote down acceptance criteria before seeing the final results (Table 1), and what was changed and why were recorded at the time for every modification to the research design. The procedure led the agent to decide, before the results existed, how large a difference had to be and how it would be tested, and the claims of these experiments were then judged against these criteria. The modeling tools alone do not provide this step, which corresponds to the kind of failure described above, in which a difference was reported without a check against random variation.

Earlier studies disagree on how much the harness matters. Ríos-García et al. (2026) found that the harness explained far less of agent behavior than the language model did; Wiedmann et al. (2026) found that dedicated verification tools changed agent behavior while verification prompts did not; Xia and Wang (2026) found that a procedure accepting only claims with recorded evidence reduced the share of unsupported claims from approximately 13.3% to 4.9%. Our experiments add a result for hydrological model development: claims that do not hold are six times as common under the general harness as under the general harness plus the HydroMind modeling tools and the research procedure. Our results also suggest a way to reconcile the earlier findings: adding tools alone changed little in how claims were judged, whereas a procedure that specifies, before the results are seen, what kind of difference can be stated as a claim changed it markedly.

This result has the following implications for researchers who use or read hydrological studies conducted by agents. Under the general harness, fewer than half of the claims have statistical support, and approximately three in ten do not hold on recheck; such claims must be recomputed from the predictions before they can be used. Without recomputing, readers can check three items from the study's own records: whether the code and settings for each comparison are on record; whether each stated difference is accompanied by the study's own statistical test, confidence interval, or decision rule; and whether acceptance criteria were written down before the final results were produced.

The experiments used only one language model; conducting the same experiments with other language models would indicate whether the effect of the harness depends on the model. The four questions studied all involve hydrological model development; adding more questions, including questions from other areas of hydrology, would show how far these results extend.

## 6 Conclusions

When a general-purpose agent makes the research decisions itself in hydrological model development, how many of its claims hold and whether readers can check them depend on its harness. The same language model studied four questions under three harnesses and submitted 84 claims. With the general harness alone, 30% of the claims did not hold on recheck, and all of these claims stated as definite a difference that could not be distinguished from random variation. With the HydroMind modeling tools added, the code and settings behind every claim were on record and the share of claims that do not hold fell to 20%, but the share of claims with the experiment's own statistical support barely changed (48% and 53%). With the HydroMind research procedure also added, the agent wrote down acceptance criteria before seeing the final results, the share of claims with statistical support rose to 83%, and the share of claims that do not hold fell to 5%. Hydrological studies that use agents therefore need, in addition to modeling tools, a requirement that the agent decide before seeing the results what kind of difference can be stated as a claim.

## Open Research

The CAMELS-US data used in this study are publicly available (Newman et al., 2015; Addor et al., 2017). HydroMind is distributed through the Python Package Index (PyPI, https://pypi.org/project/hydromind/) under the PolyForm Strict 1.0.0 license, and a license is required to use it. After registering an account at https://hydromind.funzzz.top, users can obtain a free license by sending the account name or registered email address to 1831372118fz@gmail.com; licenses can also be purchased on the same website. The HydroMind documentation is available on the same website.

## Conflict of Interest

The authors developed HydroMind, the software evaluated in this study. HydroMind licenses can be requested free of charge by email; the HydroMind website also offers licenses at a price of CNY 0.99, a fee charged solely to prevent misuse of the website's resources. The authors declare no other conflicts of interest.